\documentclass[]{aa}
\usepackage{psfig}
\begin{document}

\title{Photometry of SN 2002bo with template image subtraction}

\author{Szab\'o, Gy. M.\inst{1,3},
S\'arneczky, K.\inst{2,3}, Vink\'o, J.\inst{2,3},
Cs\'ak, B.\inst{1,3},
M\'esz\'aros, Sz.\inst{2,3}, Sz\'ekely, P.\inst{1,3},
Bebesi, Zs.\inst{3}}
\institute{Department of Experimental Physics \& Astronomical Observatory,
University of Szeged,
H-6720 Szeged, D\'om t\'er 9., Hungary \and
Department of Optics \& Quantum Electronics \& Astronomical
Observatory, University of Szeged,
POB 406, H-6701 Szeged, Hungary \and
Guest Observer at Piszk\'estet\H o Station, Konkoly Observatory, Hungary}
\titlerunning{Photometry of SN 2002bo}
\authorrunning{Szab\'o et al.}
\offprints{szgy@mcse.hu}
\date{}

\abstract{
$VRI$ photometry of the type Ia supernova 2002bo is presented. 
This SN exploded in a dusty region of the
host galaxy NGC 3190, thus, subtraction of a template frame was necessary
to obtain reliable photometry. We used a template frame of NGC 3190
taken during the course of our galaxy imaging project, fortunately, just a
few days before SN 2002bo was discovered. The aim of
this project is to collect template frames of nearby galaxies that are
potential hosts of bright SNe. Subtraction of
pre-SN images helped us to exclude the background light contamination of
the host galaxy. The maximum occurred at JD 2452346, with maximal V
brightness of 13\fm58.  MLCS analysis led to $T_0(B)$=JD 2452346.1
$\pm 0.8$ (fiducial B-maximum), $E(B-V)$=0\fm24$\pm$0\fm02,
$\mu_0=32.46\pm0.06$, $\Delta=-0.14\pm0.04$. $E(B-V)$=0\fm24(2) indicates
a significant extinction in the host galaxy as the galactic reddening is
negligible toward NGC 3190. The accepted value of $\Delta$ indicates that SN
2002bo was a
slightly overluminous SN by about 0\fm14 relative to fiducial SN Type Ia.
The distance turned out to be 31.0$\pm$3 Mpc. 
In addition, the heavily obscured SN 2002cv was also detected on the I frame
taken on JD 2452434 (June 8, 2002), and a
variable star is found in the field, very close to the host galaxy.}

\maketitle

\section{Introduction}
The bright supernova SN 2002bo in NGC 3190 (belonging to the Leo III group
LGG 194) was
spectroscopically discovered by Kawakita et al. (2002) with GAO 0.65
telescope on March 9.6 UT. A low resolution spectrum (ranged 380-750 nm)
showed a highly reddened Ia type supernova, about two weeks before maximum
light. The expansion velocity was measured as 19000 km/s. Another
spectrum (range 400-780 nm, resolution 2.5 nm) obtained on March 10.04 UT
with the Asiago 1.82 telescope, showed strong P-Cygni profiles of Si II 635
and 596 nm. The velocity measured from Si II absorption was 17700 km/s
(Benetti et al., 2002). No object brighter than 19.5 was found at the
position of the SN on CCD frames collected on Mar. 4.0 UT (S\'arneczky \&
Bebesi, 2002).

Matheson et al. (2002) reported an expansion velocity of 17900 km/s
on March 10.25 UT. Strong, narrow Na D absorption at the velocity of
the host galaxy showed that there is probably higher reddening
caused by the interstellar
matter inside NGC 3190. The galactic reddening toward
NGC 3190 is only $E(B-V)$=0\fm025, according to Schlegel et al., (1998).
Several titanium absorption lines in the spectrum
(reported by Matheson et al., 2002) suggested that this may be a subluminous
event similar to SN 1999by.

\begin{figure}
\begin{center}
\leavevmode
\psfig{figure=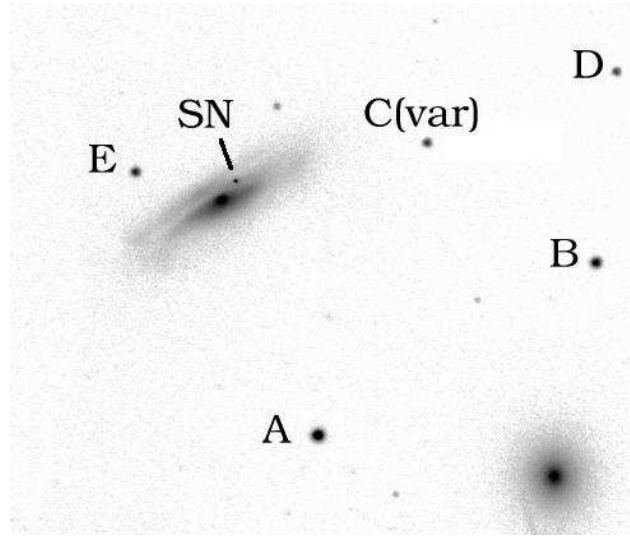,width=8.5cm}
\caption{SN 2002bo in NGC 3190 with the comparison stars and
field variable C.}
\end{center}
\end{figure}

The possible underluminosity is confirmed by Chornock et al. (2002)
from unusually strong Si~II (585 nm) absorption. With the Shane 3-m reflector
they measured an expansion velocity of about 16000 km/s from the Si~II (635.5 nm)
line on March 11.0 UT.

The host galaxy NGC 3190 (a type SA(s)a LINER)
has an integrated B magnitude 11\fm31 corrected for galactic absorption.
Its radial velocity in the optical band is +1271$\pm$14 km/s,
which corresponds to a kinematical distance modulus $\mu=-31\fm31$ assuming
$H_0=65$ km s$^{-1}$ Mpc$^{-1}$. The $D_{25}$ diameter is 4.3 arcsec,
$\epsilon_{25}$ ellipticity is 0.58 with position angle 125.
The 3-dimensional central velocity dispersion is $\sigma_v=169\pm11$
km  s$^{-1}$ according to H\'eraudeau \& Simien (1998).
SN 2002bo appeared on the edge of a dust lane in
NGC 3190 (Fig. 1.), and reached a maximum visual brightness of about
13\fm6 by mid March, 2002.

Distance modulus to NGC 3190 is not concordant in the literature.
According to H\'eraudeau \& Simien (1998), the dynamical distance modulus is
$\mu_0=31\fm31$ ($H_0=75$ km s$^{-1}$ Mpc$^{-1}$),
while based on the Virgocentric-fall corrected radial velocity,
van Driel et al. (2001) gives only 17 Mpc distance, $\mu_0=31\fm15$ ($H_0=
65$ km s$^{-1}$ Mpc$^{-1}$). However, the average radial velocity of the
Leo III group is about 1400 km/s, thus, the distance to its members may be 
more than discussed above. Based on the
Tully-Fisher relation, de Vaucouleurs (1991) gives $\mu_{TF}=31\fm61$.
Most recently, the surface brightness fluctuation method led to
a significantly higher distance modulus of
$\mu_{SBF}=32\fm38\pm0\fm15$, indicating 30 Mpc distance to NGC 3190
(Tonry et al., 2001).

\section{               Observations and data reduction}

\begin{figure}
\begin{center}
\leavevmode
\psfig{figure=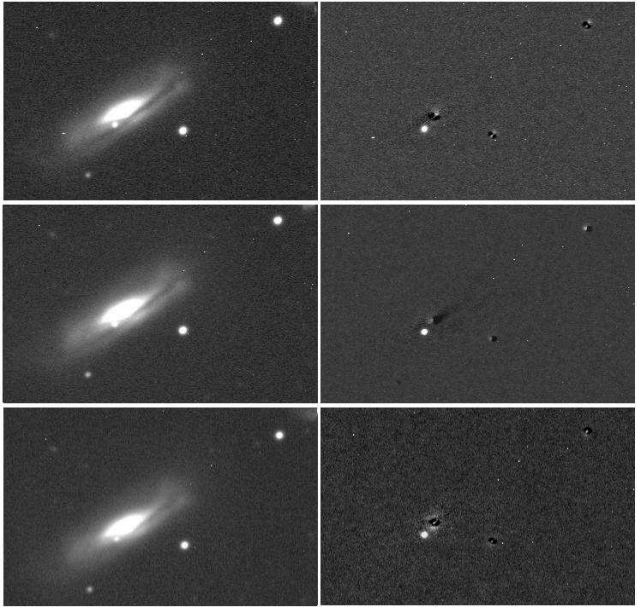,width=8.5cm}
\caption{Observed $V$, $R$ and $I$ images of SN 2002bo on 16th March (left panel),
and the residual images after template subtraction (right panel).}
\end{center}
\end{figure}

We have observed SN 2002bo on 11 nights at three observing sites. Six points
before light maximum and a post-maximum observation in June were measured
with the 60/90/180 Schmidt-telescope of Konkoly Observatory equipped
with a Photometrics AT200 CCD (1\farcs00 scale, RDN=14, Vink\'o et al., 2003).
This telescope was also used for our host-galaxy imaging project
which aims at collecting $BVRI$ frames
of the closest galaxies to be used as templates for photometry of future SNe.

Most of the points were measured using the 28cm Schmidt-Cassegrain telescope
of Szeged Observatory equipped with an SBIG ST7E CCD. This 765$\times$510 camera
gives 1\farcs0/pixel resolution, the pixels are rectangular. Despite the
unfavorable location of this telescope (in the very center of the city of
Szeged), $V$, $R$ and $I$ observations could be made with satisfactory accuracy
using integrated expositions of 10--20$\times$1 minute images.  All the
observations between JD 2452350 and JD 2452389 were made with this
equipment.

The transformation equations for Szeged Observatory and Konkoly Observatory,
respectively, were applied as follows.

\bigskip

$V=v-0.21(8)\cdot (V-R)+\kappa_V$

$(V-R)=0.83(2)\cdot (v-r)+\kappa_{V-R}$

$(V-I)=0.88(9)\cdot (v-i)+\kappa_{V-I}$
\bigskip

$V=v+0.09(3)\cdot (V-R)+\kappa_V$

$(V-R)=0.98(4)\cdot (v-r)+\kappa_{V-R}$

$(V-I)=1.13(2)\cdot (v-i)+\kappa_{V-I}$

\bigskip

Local standards were measured at Konkoly Observatory on June 8th using
standard fields PG 1633+09 and PG 1323-086 of Landolt (1992).  Magnitudes
and errors accepted for comparison stars are presented in Table 1 (see Fig.1
for the designations of the stars). Uncertainties are computed by
quadratic summation of errors in individual measurements and the
transformation coefficient errors. Star C (GSC 1425 35) varies (suspected
eclipsing variable), so it was rejected from the comparison set.

\begin{table}
\caption{GSC-numbers, magnitudes and colors of local standards.}
\begin{tabular} {lllll}
\hline
Star & GSC number & $V$ & $V-R$ & $R-I$ \\
\hline
A & 1425 496 & 12.28(4)& 0.29(3)& 0.26(8)\\
B & 1425 42 & 13.57(4)& 0.45(3)& 0.50(2)\\
D & 1425 713 & 15.31(4)& 0.70(3)& 0.60(3)\\
E & 1425 629 & 14.27(4)& 0.44(3)& 0.45(2)\\
\hline
\end{tabular}
\end{table}

\begin{table}
\caption{Observation time (JD-2450000) and standard photometry
of SN 2002bo. Telescope codes: A: 60/90/180 cm Schmidt, Konkoly Obs.,
B: 28 cm Schmidt-Cassegrain, Szeged Obs. The errors in parentheses refer
to the accuracy of the relative photometry; 
see text for the systematic errors in different bands.}
\begin{tabular} {lllll}
\hline
Time &Tel. & $V$ & $R$ & $I$\\
\hline
2344.285 &A& 15.729(0.02) & 15.47(0.02) & 15.70(0.02) \\
2345.257 &A& 15.320(0.02) & 15.02(0.02) & 15.18(0.02) \\
2350.350 &B& 13.945(0.03) & 13.67(0.04) & --- \\
2365.354 &B& 13.757(0.03) & 13.62(0.04) & 13.96(0.04) \\
2367.344 &B& 13.931(0.03) & 13.91(0.04) & --- \\
2371.313 &B& 14.131(0.03) & 14.05(0.04) & 14.10(0.04) \\
2380.308 &B& 14.469(0.06) & 14.09(0.09) & 14.02(0.09) \\
2381.323 &B& 14.557(0.06) & 14.22(0.10) & 14.13(0.11) \\
2388.311 &B& 14.967(0.07) & 14.51(0.11) & 14.27(0.13) \\
2396.313 &B& 15.314(0.09) & 15.07(0.11) & ---          \\
2434.406 &A& 16.405(0.05) & 16.26(0.07) & 16.19(0.11) \\
\hline
\end{tabular}
\end{table}

SN 2002bo appeared close to the center of the host galaxy NGC3190, so the
brightness contribution of NGC 3190 must be taken into account in the
reduction. The dust lane of this edge-on spiral also passes near the
position of the supernova, causing a quite inhomogeneous background. In
similar cases, the observer usually has to wait at least a year after the
explosion, when the fireball diminishes enough to allow imaging the
background brightness contribution of the host galaxy.

\subsection{            Subtraction of the template}

In order to get acceptable photometry, the subtraction of a template image
of NGC 3190 was necessary. Fortunately, this image was taken on March 4.0
UT, 2002 during the course of our galaxy imaging project. The aim of this
image subtraction was to smooth out the background of the SN, so that the
scraggy surface of NGC 3190 does not disturb the reduction. For the proper
subtraction, SN- and template images must have the same geometry, and the
same flux scale above the zeroed background. Thus, the template and SN
frames were registered and rescaled before the template subtraction.

The geometrical transformations of the template frames consisted of a
stretch and a rotation, so higher-order distortions caused by the image
curvature were neglected. We used the $imexamine$ task in
IRAF\footnote{$IRAF$ is distributed by NOAO which is operated by the
Association of Universities for Research in Astronomy (AURA) Inc. under
cooperative agreement with the National Science Foundation.} to make
accurate astrometry of the comparison stars and the SN. The linear
geometrical transformations were calculated by the $geotran$ and $imarith$
tasks. The fixed point of the transformations was the supernova itself, so
higher-order geometrical distortions were almost negligible in the close
surroundings.

As a second step, the backgrounds were zeroed on each image. The
subtracted sky values were measured manually on the individual images.
Fluxes of comparison stars were calculated by $imexamine$. The intensities
of the SN image were rescaled so that the total fluxes (the summarized ADU
values) of comparison stars A, B, D, E were the same as on the template
frame.

It should be noted that the filter characteristics on Konkoly and
Szeged sites are quite different. One could therefore think that the
subtraction of the nominally adequate images could result in considerable
residuals, as the two filters are centered on slightly different bands.
Fortunatelly, even in this case the subtraction offers much better
circumstances for reduction, as the overall spectrum of the host galaxy
(disk region) is quite smooth. After rescaling the fluxes of the SN image,
we simply subtract ``something similar'' to the host galaxy but {\it without}
the SN. Perfect subtraction should therefore result in a clear image where
nothing but the supernova is visible. Its flux is scaled onto the template
system, but this does not distort the photometry.

After the template subtraction the host galaxy together with the foreground
stars should vanish completely. In practice, however, the PSF slightly
varies from image to image, causing varying structures and weights of the
profile veils. Therefore, the PSF of the observed image was measured near
the SN, and the template image was blurred with a gaussian in order to have
approximately the same PSF as the SN frames.

The result is presented in Fig. 2 where the observed and the residual images
are compared. Fainter stars have vanished completely, though the higher-order
component of the veils of the brighter stars remained on the frame. Though
the galaxy is not perfectly subtracted, the background around the SN is much
smoother and it could be removed much better during the photometry.

The final step was the application of a standard aperture photometry. The
flux of the comparison stars were measured on the observed frames after
geometric transformations and intensity scaling, but before template
subtraction. The SN was measured after the subtraction. Nominal errors of
the relative photometry were calculated by $IRAF$, analysing the scatter of
the background near the SN and the detected flux of the SN itself.

The random errors of the differential photometry were estimated as 
$\sigma_{rnd}^2=\sigma_p^2 + \sigma_{c}^2 + \sigma_b^2$, where 
$\sigma_p$ is the
error of the photometry caused by the photon noise, $\sigma_c$ is 
caused by the color-dependent part of the transformations (e.g.
$\Delta \epsilon \cdot (V-R) + \Delta (V-R) \cdot \epsilon)$), and
$\sigma_b$ is the error caused by the template subtraction, namely the 
resultant inhomogeneities in the residual
background, typically 2-6\%, depending on the quality of the observed
images. \footnote{In our experience, the residual sky never exceeds 5\% of
the value of the originally observed sky.
So we estimated $\sigma_b(\%) \approx { SKY\cdot 0.05 \over FLUX}$,
where SKY and FLUX refer to the sky value inside the aperture and the
measured brightness of the SN, respectively.}

Systematic errors of the standard photometry come from the measurement of
the local standard stars. Those were estimated as $\sigma_V=0\fm04$,
$\sigma_R=0.03$, $\sigma_i=0.04$. These systematic errors do not influence 
the shape of the light curve
or the $\Delta$ and $T$ parameters in the MLCS-modelling. However, they
are present the distance modulus and the extinction. Therefore we added
$\sigma_{\rm sys} \approx 0\fm04$ to the fit errors of $\mu_0$ and $E(B-V)$.

\section{                            Results}

\begin{figure}
\begin{center}
\leavevmode
\psfig{figure=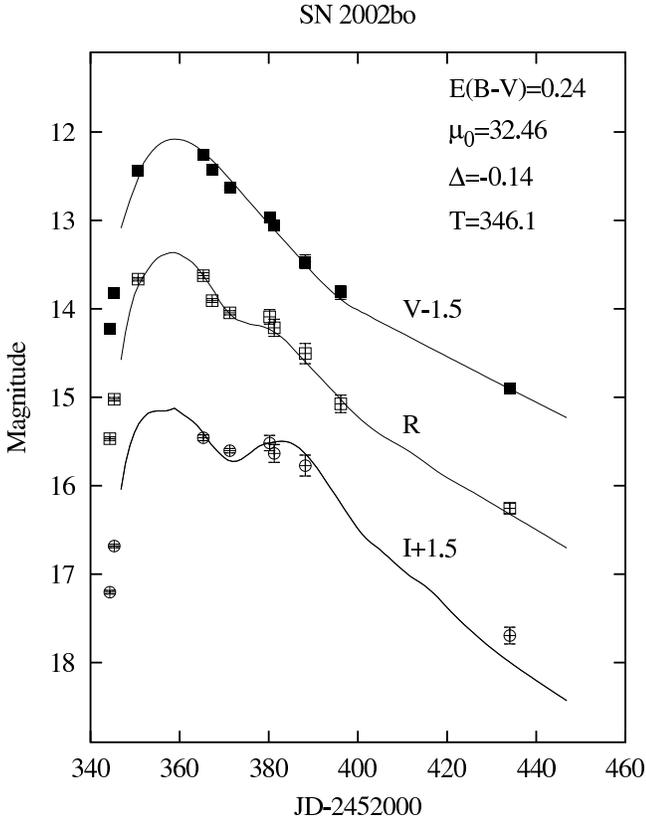,width=8.5cm}
\caption{Observed $V$, $R$ and $I$ light curves of SN 2002bo with
the calculated MLCS light curve fit.}
\end{center}
\end{figure}

The calibrated standard magnitudes of SN~2002bo are listed in Table~2. The
estimated errors of each point are given in parentheses. 
The $VRI$ light curves were analyzed by the
Multi-Color Light Curve Shape (MLCS) methods of Riess et al. (1996, 1998).

The timescale of the observations was divided by $1+z=1.0042$ to remove
the effect of time dilation. The fitting was computed simultaneously for all
light curves (see Vink\'o et al., 2001 for the description of the fitting
method). The $VRI$ light curves of the SN with the resulted MLCS model 
curves are presented in Fig. 3. The two
earliest points before the maximum could not be fitted (although
they are presented in Fig.3), because the fiducial
curves do not cover the early ascending branch before $-10$ days. The
last point measured in the $I$ band seems to be too bright. This
point was, however, included in the fitted dataset, as the individual $I$
images are of a good quality.

The residuals of the fitting are shown in Fig. 4. All points could be fitted
within $\pm 0.2$ mag, the overall standard deviation is $\sigma = 0\fm115$.

The minimum of $\chi ^2$ was found at the following parameters.
$T_0(B)$=JD 2452346.1
$\pm 0.8$ (fiducial B-maximum), $E(B-V)$=0\fm24$\pm$0\fm02,
$\mu_0=32.46\pm0.06$, $\Delta=-0.14\pm0.04$.
The given confidence intervals
correspond to 1-dimensional $\chi ^2$ tests of each individual parameter with
a significance level of 90\%.

\begin{figure}
\begin{center}
\leavevmode
\psfig{figure=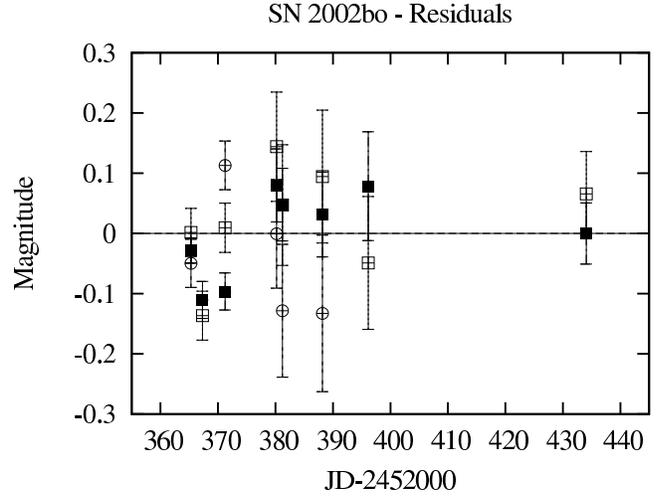,width=8.5cm}
\caption{Fitting residuals of the MLCS method. Symbols are the same as in
Fig. 3.}
\end{center}
\end{figure}

\begin{figure}
\leavevmode
\psfig{figure=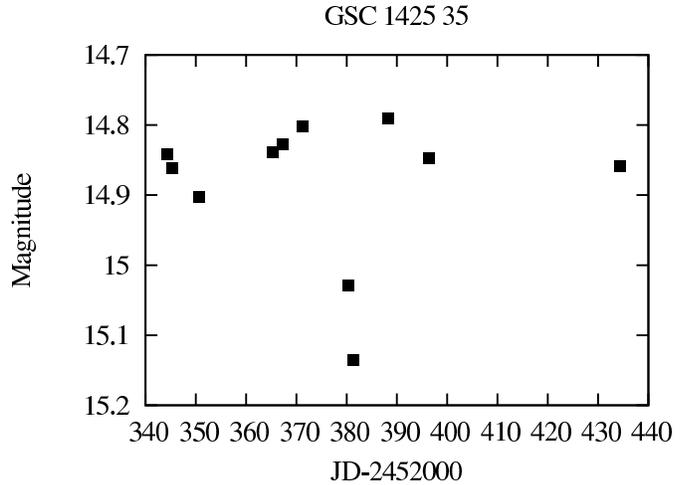,width=8.8cm}
\caption{V-band light curve of variable star C = GSC 1425 35.}
\end{figure}

The parameter $\Delta$ suggests a slight overluminous SN.
It is an interesting result, because early spectroscopy
suggested a possible underluminous supernova (from the ratio
of depths of the 5850 and 6355 \AA~Si~II lines, see Sect.1.).
On the other hand, underluminosity would result in
a precipitous decline rate. In our photometry, decline rates were found to
be about normal or slightly steep, especially for the well-fitted $V$ and $R$
curves. The contradiction that early spectroscopy indicated an
underluminous explosion, but later the photometry has shown a normal
supernova, is an interesting subject for later studies.

The MLCS analysis resulted in a new distance modulus to the host galaxy,
$\mu_0 = 32.46 \pm 0.06$ mag (adopting $M_V = -19.46$ for the fiducial SN curve).
The real error of $\mu_0$ is probably increased by the
errors of local standard magnitudes. Therefore, we suggest
$\mu_0 = 32.46 \pm 0.10$ assuming that absolute magnitudes of
local standards are measured more accurately than $0\fm08$.

Comparing the resulting color excess to the reddening due to Milky Way dust
($E(B-V) = $0\fm025 mag, Schlegel et al. 1998), it can be seen that the
host galaxy ISM highly
reddens SN~2002bo. This result is concordant with the spectroscopic results
detailed in Sect. 1. Adopting E($B-V$)=0\fm24 mag, the total absorption
in $V$, $R$ and $I$ bands is $A_V=0\fm77$, $A_R=0\fm62$, $A_I=0\fm43$ using the
reddening law given by Schlegel et al. (1998).

\subsection{             Variable star in the NGC 3190 field}

While selecting the appropriate comparison stars, differential magnitudes of
the candidates (A, B, C, D, E stars in Fig. 1.) were calculated. All stars
showed constant light curves, but star C (GSC 1425 35) seemed to vary. This
star was excluded from the comparison star set, and its light variation was
measured with respect to stars A, B, D and E.

The undersampled $V$-light curve (Fig. 5.)
of star C shows at least 0\fm35 mag variation with a narrow
primary minimum at about JD 2452379 and probable secondary minimum at
about JD 2452356$\pm$6. The general appearance suggests an eclipsing variable.
The continuous variation might be due to $\beta$ Lyrae type, but a
longer period may result in an Algol type light curve
with considerable reflection
effect. Its color indexes on JD 2434.406 UT were $V-R$=0\fm52(3),
$R-I$=0\fm47(2), which are not typical for Algol variables.
However, based on few selected data points,
no firm conclusions can be drawn either on
the variability type or the light curve characteristics.

\subsection{                Detection of SN 2002cv}

\begin{figure}
\leavevmode
\psfig{figure=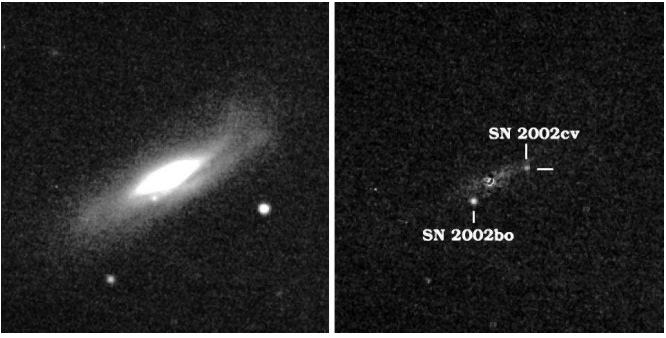,width=8.8cm}
\caption{$I$ filtered observation of SN 2002bo and SN 2002cv on 8th June.}
\end{figure}

The heavily obscured supernova SN 2002cv was discovered by Di Paola et
al. (2003) on the 13th May. This is probably the most reddened supernova
ever observed, its $A_V$ absorption is 7\fm8$\pm$1\fm0. 

We also detected this object on JD 2452434
(June 8, 2002), about 19 days after its maximum (Fig. 6.). Despite the
unfavourable sky conditions  and high airmass, it is visible
in the $I$-filtered image, taken with a total exposure time of 5 minutes.
The template subtraction allowed us to measure its brightness, which turned
out to be 17\fm2$\pm$0\fm3. This value is in a very good aggreement with the
light curve taken by Di Paola et al (2003). As far as we know, we have 
measured the last photometric point before the conjunction. Examining the 
$R$ and $V$ images, no firm evidence for the presence of 
SN 2002cv is found because of the high reddening.

\section{                      Conclusions}

\begin{enumerate}

\item $VRI$ photometry of SN 2002bo is presented starting from 12 days before
maximum light and extending up to 80 days past maximum. The maximum in the
$B$ band occurred
at about JD 2452346.1. In $V$ the brightness reached 13\fm58 mag in maximum.
Thanks to its detection in the $I$ band, a late photometric
point is presented for SN 2002cv at JD 2452424.4, $I$=17\fm2 $\pm$ 0\fm3.
A variable star is found in the field that is very close to host galaxy.

\item The reddening was estimated by applying the MLCS
method for all light curves. This yielded $E(B-V)$=0\fm24(2) indicating
significant extinction in host galaxy, while absorption caused by
the the host galaxy is $A_V$=0\fm77 using a standard reddening law. 
Spectroscopic observations confirms the presence of significant absorption 
by the ISM in NGC 3190.

The MLCS method assumes that the standard Galactic reddening law is valid in
the SN host galaxy as well. It is worth noting that the observed $VRI$ light
curves could be fitted quite well when most of the reddening is due to
the ISM in the host galaxy. This supports the reliability of the assumption
made above.

\item  The light curve analysis indicated that SN 2002bo was a slightly
overluminous SN by about 0\fm14 relative to fiducial SN Type Ia. This
is supported by the shape and the decline rate of the light curves, but
is in contradiction with early spectroscopic data, where stronger Si II 
absorption features may have indicated a slightly subluminous explosion.

\item The distance of SN 2002bo inferred from the MLCS method was found to be
31.0$\pm$3 Mpc. This distance is about 50\% higher than early dynamical
and Tully-Fisher distances, but agrees fairly
well with the most recent SBF determinations. These findings are
concordant with earlier results that SBF and SN Ia distances both
support the long extragalactic distance scale (de Vaucouleurs, 1983, 
Ajhar et al, 2001).

\end{enumerate}

\begin{acknowledgements}
This work has been supported by the Hungarian OTKA Grants T034315 and
FKFP Grant 0010/2001. The kind hospitality of the Konkoly
Observatory and their provision of telescope time is gratefully
acknowledged. We thank our referee T. Matheson for his mindful reading and
suggestions.

\end{acknowledgements}


\thebibliography{}

\bibitem []{} Ajhar, E.A., Tonry, J.L., Blakeslee, J.P., Riess, A.G.,
Schmidt, B.P., 2001, ApJ 559, 584

\bibitem[]{} Benetti, S., Altavilla, G., Pastorello, A.,
et al., 2002, IAUC 7848

\bibitem[]{} Chornock, R., Li, W.D., Filippenko, A.V., 2002, IAUC 7851

\bibitem []{} Di Paola,
A., Larionov, V., Arkharov, A. et al. 2002, A\&A 393, L21

\bibitem[]{} van Driel, W., Marcum, P., Gallagher, J.S., 2001, A\&A 378, 370 

\bibitem[]{} Kawakita, H., Kinugasa, K., Ayani, K., Yamaoka,
H., 2002, IAUC 7848

\bibitem[]{}  Landolt A.U., 1992, AJ 104, 340

\bibitem[]{} Matheson, T., Jha, S., Challis, P., Kirshner, R., 2002, IAUC
7849

\bibitem[]{} H\'eraudeau, Ph., Simien, F., 1998, A\&AS 133, 317

\bibitem[]{} Riess, A.G., Press, W.H., Kirshner, R.P., 1996, ApJ 473, 88

\bibitem[]{} Riess, A.G., Filippenko, A.V., Challis P., et al., 1998, AJ,
116, 1009

\bibitem[]{} S\'arneczky, K., Bebesi, Zs., 2002, IAUC 7863

\bibitem[]{} Schlegel, D.J., Finkbeiner, D.P., Davis, M., 1998, ApJ 500, 525.

\bibitem[]{} Tonry, J.L., Dressler, A., Blakeslee, J.P., et al., 2001, ApJ
546, 681

\bibitem[]{} de Vaucouleurs, G., 1983, ApJ 268, 468

\bibitem[]{} de Vaucouleurs, G., de Vaucouleurs, H., Corwin, H.G., 1991,
Third Reference Catalog of Bright Galaxies, Springer, New York

\bibitem[]{} Vink\'o, J., B\'ir\'o, I.B., Cs\'ak, B., et al., 2003, A\&A 397,
115

\bibitem[]{} Vink\'o, J., Cs\'ak, B., Csizmadia, Sz., et al., 2001, A\&A 372,
824
\end{document}